\documentclass[
    ,final            
  ]
  {aipproc}

\layoutstyle{6x9}

\begin{document}
%
\newbox\hdbox%
\newcount\hdrows%
\newcount\multispancount%
\newcount\ncase%
\newcount\ncols
\newcount\nrows%
\newcount\nspan%
\newcount\ntemp%
\newdimen\hdsize%
\newdimen\newhdsize%
\newdimen\parasize%
\newdimen\spreadwidth%
\newdimen\thicksize%
\newdimen\thinsize%
\newdimen\tablewidth%
\newif\ifcentertables%
\newif\ifendsize%
\newif\iffirstrow%
\newif\iftableinfo%
\newtoks\dbt%
\newtoks\hdtks%
\newtoks\savetks%
\newtoks\tableLETtokens%
\newtoks\tabletokens%
\newtoks\widthspec%
%
%
\immediate\write15{%
CP SMSG GJMSINK TEXTABLE --> TABLE MACROS V. 851121 JOB = \jobname%
}%
%
%
\tableinfotrue%
\catcode`\@=11
\def\out#1{\immediate\write16{#1}}
%
%
\def\tstrut{\vrule height3.1ex depth1.2ex width0pt}%
\def\and{\char`\&}
\def\tablerule{\noalign{\hrule height\thinsize depth0pt}}%
\thicksize=1.5pt
\thinsize=0.6pt
\def\thickrule{\noalign{\hrule height\thicksize depth0pt}}%
\def\hrulefill{\leaders\hrule\hfill}%
\def\bigrulefill{\leaders\hrule height\thicksize depth0pt \hfill}%
\def\ctr#1{\hfil\ #1\hfil}%
\def\altctr#1{\hfil #1\hfil}%
\def\vctr#1{\hfil\vbox to0pt{\vss\hbox{#1}\vss}\hfil}%
%
%
\tablewidth=-\maxdimen%
\spreadwidth=-\maxdimen%
\def\tabskipglue{0pt plus 1fil minus 1fil}%
%
%
\centertablestrue%
\def\centeredtables{%
   \centertablestrue%
}%
\def\noncenteredtables{%
   \centertablesfalse%
}%
%
%
\parasize=4in%
\long\def\para#1{
   {%
      \vtop{%
         \hsize=\parasize%
         \baselineskip14pt%
         \lineskip1pt%
         \lineskiplimit1pt%
         \noindent #1%
         \vrule width0pt depth6pt%
      }%
   }%
}%
\gdef\ARGS{########}
\gdef\headerARGS{####}
\def\@mpersand{&}
{\catcode`\|=13
\gdef\letbarzero{\let|0}
\gdef\letbartab{\def|{&&}}%
\gdef\letvbbar{\let\vb|}%
}
{\catcode`\&=4
\def\ampskip{&\omit\hfil&}
\catcode`\&=13
\let&0
\xdef\letampskip{\def&{\ampskip}}%
\gdef\letnovbamp{\let\novb&\let\tab&}
}
\def\begintable{
   \begingroup%
   \catcode`\|=13\letbartab\letvbbar%
   \catcode`\&=13\letampskip\letnovbamp%
   \def\multispan##1{
      \omit \mscount##1%
      \multiply\mscount\tw@\advance\mscount\m@ne%
      \loop\ifnum\mscount>\@ne \sp@n\repeat%
   }
   \def\|{%
      &\omit\widevline&%
   }%
   \ruledtable
}
\long\def\ruledtable#1\endtable{%
%
%
%
   \offinterlineskip
   \tabskip 0pt
   \def\widevline{\vrule width\thicksize}
   \def\endrow{\@mpersand\omit\hfil\crnorm\@mpersand}%
   \def\crthick{\@mpersand\crnorm\thickrule\@mpersand}%
   \def\crthickneg##1{\@mpersand\crnorm\thickrule
          \noalign{{\skip0=##1\vskip-\skip0}}\@mpersand}%
   \def\crnorule{\@mpersand\crnorm\@mpersand}%
   \def\crnoruleneg##1{\@mpersand\crnorm
          \noalign{{\skip0=##1\vskip-\skip0}}\@mpersand}%
   \let\nr=\crnorule
   \def\endtable{\@mpersand\crnorm\thickrule}%
   \let\crnorm=\cr
%
%
   \edef\cr{\@mpersand\crnorm\tablerule\@mpersand}%
   \def\crneg##1{\@mpersand\crnorm\tablerule
          \noalign{{\skip0=##1\vskip-\skip0}}\@mpersand}%
   \let\ctneg=\crthickneg
   \let\nrneg=\crnoruleneg
   \the\tableLETtokens
%
%
   \tabletokens={&#1}
%
%
   \countROWS\tabletokens\into\nrows%
   \countCOLS\tabletokens\into\ncols%
%
%
   \advance\ncols by -1%
   \divide\ncols by 2%
   \advance\nrows by 1%
%
%
   \iftableinfo %
      \immediate\write16{[Nrows=\the\nrows, Ncols=\the\ncols]}%
   \fi%
%
%
   \ifcentertables
      \ifhmode \par\fi
      \hbox to \hsize{
      \hss
   \else %
      \hbox{%
   \fi
      \vbox{%
         \makePREAMBLE{\the\ncols}
         \edef\next{\preamble}
         \let\preamble=\next
         \makeTABLE{\preamble}{\tabletokens}
      }
      \ifcentertables \hss}\else }\fi
   \endgroup
   \tablewidth=-\maxdimen
   \spreadwidth=-\maxdimen
}
\def\makeTABLE#1#2{
   {
   \let\ifmath0
   \let\header0
   \let\multispan0
%
%
   \ncase=0%
   \ifdim\tablewidth>-\maxdimen \ncase=1\fi%
   \ifdim\spreadwidth>-\maxdimen \ncase=2\fi%
   \relax
%
   \ifcase\ncase %
      \widthspec={}%
   \or %
      \widthspec=\expandafter{\expandafter t\expandafter o%
                 \the\tablewidth}%
   \else %
      \widthspec=\expandafter{\expandafter s\expandafter p\expandafter r%
                 \expandafter e\expandafter a\expandafter d%
                 \the\spreadwidth}%
   \fi %
   \xdef\next{
      \halign\the\widthspec{%
      #1
      \noalign{\hrule height\thicksize depth0pt}
      \the#2\endtable
%
      }
   }
   }
   \next
}
\def\makePREAMBLE#1{
   \ncols=#1
   \begingroup
   \let\ARGS=0
   \edef\xtp{\widevline\ARGS\tabskip\tabskipglue%
   &\ctr{\ARGS}\tstrut}
   \advance\ncols by -1
   \loop
      \ifnum\ncols>0 %
      \advance\ncols by -1%
      \edef\xtp{\xtp&\vrule width\thinsize\ARGS&\ctr{\ARGS}}%
   \repeat
   \xdef\preamble{\xtp&\widevline\ARGS\tabskip0pt%
   \crnorm}
   \endgroup
}
\def\countROWS#1\into#2{
   \let\countREGISTER=#2%
   \countREGISTER=0%
   \expandafter\ROWcount\the#1\endcount%
}%
\def\ROWcount{%
   \afterassignment\subROWcount\let\next= %
}%
\def\subROWcount{%
   \ifx\next\endcount %
      \let\next=\relax%
   \else%
      \ncase=0%
      \ifx\next\cr %
         \global\advance\countREGISTER by 1%
         \ncase=0%
      \fi%
      \ifx\next\endrow %
         \global\advance\countREGISTER by 1%
         \ncase=0%
      \fi%
      \ifx\next\crthick %
         \global\advance\countREGISTER by 1%
         \ncase=0%
      \fi%
      \ifx\next\crnorule %
         \global\advance\countREGISTER by 1%
         \ncase=0%
      \fi%
      \ifx\next\crthickneg %
         \global\advance\countREGISTER by 1%
         \ncase=0%
      \fi%
      \ifx\next\crnoruleneg %
         \global\advance\countREGISTER by 1%
         \ncase=0%
      \fi%
      \ifx\next\crneg %
         \global\advance\countREGISTER by 1%
         \ncase=0%
      \fi%
      \ifx\next\header %
         \ncase=1%
      \fi%
      \relax%
      \ifcase\ncase %
         \let\next\ROWcount%
      \or %
         \let\next\argROWskip%
      \else %
      \fi%
   \fi%
   \next%
}
\def\counthdROWS#1\into#2{%
\dvr{10}%
   \let\countREGISTER=#2%
   \countREGISTER=0%
\dvr{11}%
\dvr{13}%
   \expandafter\hdROWcount\the#1\endcount%
\dvr{12}%
}%
\def\hdROWcount{%
   \afterassignment\subhdROWcount\let\next= %
}%
\def\subhdROWcount{%
   \ifx\next\endcount %
      \let\next=\relax%
   \else%
      \ncase=0%
      \ifx\next\cr %
         \global\advance\countREGISTER by 1%
         \ncase=0%
      \fi%
      \ifx\next\endrow %
         \global\advance\countREGISTER by 1%
         \ncase=0%
      \fi%
      \ifx\next\crthick %
         \global\advance\countREGISTER by 1%
         \ncase=0%
      \fi%
      \ifx\next\crnorule %
         \global\advance\countREGISTER by 1%
         \ncase=0%
      \fi%
      \ifx\next\header %
         \ncase=1%
      \fi%
\relax%
      \ifcase\ncase %
         \let\next\hdROWcount%
      \or%
         \let\next\arghdROWskip%
      \else %
      \fi%
   \fi%
   \next%
}%
{\catcode`\|=13\letbartab
\gdef\countCOLS#1\into#2{%
   \let\countREGISTER=#2%
   \global\countREGISTER=0%
   \global\multispancount=0%
   \global\firstrowtrue
   \expandafter\COLcount\the#1\endcount%
   \global\advance\countREGISTER by 3%
   \global\advance\countREGISTER by -\multispancount
}%
\gdef\COLcount{%
   \afterassignment\subCOLcount\let\next= %
}%
{\catcode`\&=13%
\gdef\subCOLcount{%
   \ifx\next\endcount %
      \let\next=\relax%
   \else%
      \ncase=0%
      \iffirstrow
         \ifx\next& %
            \global\advance\countREGISTER by 2%
            \ncase=0%
         \fi%
         \ifx\next\span %
            \global\advance\countREGISTER by 1%
            \ncase=0%
         \fi%
         \ifx\next| %
            \global\advance\countREGISTER by 2%
            \ncase=0%
         \fi
         \ifx\next\|
            \global\advance\countREGISTER by 2%
            \ncase=0%
         \fi
         \ifx\next\multispan
            \ncase=1%
            \global\advance\multispancount by 1%
         \fi
         \ifx\next\header
            \ncase=2%
         \fi
         \ifx\next\cr       \global\firstrowfalse \fi
         \ifx\next\endrow   \global\firstrowfalse \fi
         \ifx\next\crthick  \global\firstrowfalse \fi
         \ifx\next\crnorule \global\firstrowfalse \fi
         \ifx\next\crnoruleneg \global\firstrowfalse \fi
         \ifx\next\crthickneg  \global\firstrowfalse \fi
         \ifx\next\crneg       \global\firstrowfalse \fi
      \fi
\relax
      \ifcase\ncase %
         \let\next\COLcount%
      \or %
         \let\next\spancount%
      \or %
         \let\next\argCOLskip%
      \else %
      \fi %
   \fi%
   \next%
}%
\gdef\argROWskip#1{%
   \let\next\ROWcount \next%
}
\gdef\arghdROWskip#1{%
   \let\next\ROWcount \next%
}
\gdef\argCOLskip#1{%
   \let\next\COLcount \next%
}
}
}
\def\spancount#1{
   \nspan=#1\multiply\nspan by 2\advance\nspan by -1%
   \global\advance \countREGISTER by \nspan
   \let\next\COLcount \next}%
\def\dvr#1{\relax}%
\def\header#1{%
\dvr{1}{\let\cr=\@mpersand%
\hdtks={#1}%
\counthdROWS\hdtks\into\hdrows%
\advance\hdrows by 1%
\ifnum\hdrows=0 \hdrows=1 \fi%
\dvr{5}\makehdPREAMBLE{\the\hdrows}%
\dvr{6}\getHDdimen{#1}%
{\parindent=0pt\hsize=\hdsize{\let\ifmath0%
\xdef\next{\valign{\headerpreamble #1\crnorm}}}\dvr{7}\next\dvr{8}%
}%
}\dvr{2}}
\def\makehdPREAMBLE#1{
\dvr{3}%
\hdrows=#1
{
\let\headerARGS=0%
\let\cr=\crnorm%
\edef\xtp{\vfil\hfil\hbox{\headerARGS}\hfil\vfil}%
\advance\hdrows by -1
\loop
\ifnum\hdrows>0%
\advance\hdrows by -1%
\edef\xtp{\xtp&\vfil\hfil\hbox{\headerARGS}\hfil\vfil}%
\repeat%
\xdef\headerpreamble{\xtp\crcr}%
}
\dvr{4}}
\def\getHDdimen#1{%
\hdsize=0pt%
\getsize#1\cr\end\cr%
}
\def\getsize#1\cr{%
\endsizefalse\savetks={#1}%
\expandafter\lookend\the\savetks\cr%
\relax \ifendsize \let\next\relax \else%
\setbox\hdbox=\hbox{#1}\newhdsize=1.0\wd\hdbox%
\ifdim\newhdsize>\hdsize \hdsize=\newhdsize \fi%
\let\next\getsize \fi%
\next%
}%
\def\lookend{\afterassignment\sublookend\let\looknext= }%
\def\sublookend{\relax%
\ifx\looknext\cr %
\let\looknext\relax \else %
   \relax
   \ifx\looknext\end \global\endsizetrue \fi%
   \let\looknext=\lookend%
    \fi \looknext%
}%
%
%
\def\tablelet#1{%
   \tableLETtokens=\expandafter{\the\tableLETtokens #1}%
}%
\catcode`\@=12

\title{ High energy hadron production Monte Carlos 
\footnote{ presented at Hadronic Shower simulation workshop, FERMILAB
Sept. 6-8, 2006}}

\classification{<12.40.Nn, 13.85.Ni, 13.85.Tp>}
\keywords      {<Monte Carlo models, Inclusive hadron production, Dual
Parton model, Quantum Molecular Dynamics model>}

\author{ J.Ranft}{ address={Siegen University, Germany} }

\begin{abstract}
We discuss here Quantum molecular dynamics models { (QMD)}
and Dual Parton Models ({ DPM} and { QGSM}).
We compare { RHIC} data to {DPM}--models and we present 
a  (Cosmic ray  oriented)
{model comparison}.

\end{abstract}

\maketitle



\section{Quantum molecular dynamics models
      { (QMD)} }
The emphasis of this contribution is to high energies, therefore, 
{ { QMD} non--relativistic models ($E < 2 AGeV$) are not treated here.   }

{ The first relativistic  QMD model was {  RQMD} \cite{Sorge89a}.
This model is no longer supported
since the year 2000 ,RQMD is used in FLUKA for 
A--A collisions below 5AGeV.  } 

{ A second relativistic model is { UrQMD} \cite{Bass98}, it is 
used in CORSIKA Cosmic Ray cascade code below 80AGeV.   }

{Let me mention some efforts within the  FLUKA collaboration
which will not be treated here:
{   F.Cerutti et al.} add (approximate) 
energy conservation ,
evaporation, and residual nuclei to RQMD \cite{Cerutti04}.  }
{ { M.V.Garzelli  et al.} construct a low enery QMD for
A--A collisions in FLUKA \cite{Garzelli06}.}
{ There are efforts in  { Milano and Houston } to construct a fully 
relativistic
model for A--A collisions to be inserted into FLUKA.}

{A relativistic QMD model is a Lorentz invariant cascade 
{ (molecular dynamics)} 
with nucleons of both nuclei and all produced hadrons as participants.   }
Properties of such models are:
{(i)A formation zone cascade of all produced hadrons.  }
{(ii) Elementary interactions used in the models include:
(1) { h + h $\longrightarrow$} resonance;  {
resonance + resonance} and 
resonance decay (similar to HADRIN in FLUKA), 
(2) high enery: { h + h $\longrightarrow$} hadronic chain;
 { 2 hadronic chains} and Lund like chain fragmentation,
(3)  { chain fusion} (called formation of color ropes) in RQMD,
(4)  { empirical} parametrization of all  {
cross sections},
(5) pQCD description of  { hard collisions} (UrQMD).}

\subsection{
{   The kinematics of the UrQMD model  
   }}

The model is fully described in \cite{Bass98}.
It is based on the  covariant propagation of all hadrons
 on classical trajectories in
combination with stochastic binary scatterings, color string
formation and resonance decay. 
It includes the{ Monte Carlo solution} of a
 set of coupled partial integro-differential equations. 
Each nucleon  is represented by a coherent state ($\hbar,c =1$)
\begin{equation}
\label{gaussians}
\phi_i (\vec{x}; \vec{q}_i,\vec{p}_i,t) =
\left({\frac{2 }{L\pi}}\right)^{3/4}\, \exp \left\{
-\frac{2}{L}(\vec{x}-\vec{q}_i(t))^2 +  {\rm i} \vec{p}_i(t)
\vec{x} \right\}
\end{equation}
which is characterized by { 6 time-dependent parameters,
$\vec{q}_{i}$ and $\vec{p}_{i}$}.
 $L$, (the extension of the wave packet in
coordinate space) is fixed.
The total $n$-body wave function is a 
product of coherent states (\ref{gaussians})
$\Phi = \prod_i \phi_i (\vec{x}, \vec{q_i}, \vec{p_i}, t)$.
The Hamiltonian $H$ of the system
contains a kinetic term and mutual interactions $V_{ij}$
{ 
($H = \sum_i T_i + {\ \frac{1 }{2}} \sum_{ij} V_{ij}$)}.
This yields an Euler-Lagrange equation for each parameter.
{ 
\begin{equation}\label{hamiltoneq}
\dot{\vec{p}}_i = - \frac{\partial \langle H \rangle}{\partial \vec{q}_i}
\quad {\rm and} \quad
\dot{\vec{q}}_i = \frac{\partial \langle H \rangle}{\partial \vec{p}_i} \, .
\end{equation}
%
%
%
}
These are the time evolution equations which are solved numerically.
{
The UrQMD Hamiltonian contains:} $E_{kin}, E^{Sk2}_{jk}, E^{Sk3}_{jkl}, 
E^{Yukawa}_{jk}, E^{Coulomb}_{jk}, E^{Pauli}_{jk}$.
{

Please note: As one sees from the Hamiltonian UrQMD is  not  really a Lorentz invariant molecular
dynamics.} Therefore, the results of the model might depend strongly on
the reference frame in which the calculation is done. 
To minimize this frame dependence the authors \cite{Bass98} use a
frame--independent definition of the cross sections (via using the
impact parameters in the two--particle rest frames). 
{
They give as example for this minimization in S--S collisions}:
The multiplicities and collision numbers 
vary  only by less than 3\%
between the lab and CM frames.

{
Let us mention, that RQMD} has a  manifestly Lorentz invariant 
eq. of motion.
{
Using 4--vectors for positions and momenta},
each particle carries its own time.
{
The 2N additional degrees of freedom are fixed by 2N constraints}:
{
The N mass shell constraints:} $H_i =  p_i^2 - m_i^2 - V_i = 0$,
 ($V_i$: quasi potential) and
{
(N-1) time fixations}.
The 2Nth constraint: A relation of times of particles to the
evolution parameter $\tau$.

{ 
Projectile or target nucleus} are  modeled according to  a
Fermi-gas ansatz.
The centroids
of the Gaussians are randomly distributed within a sphere with the 
radius $R(A)$,
{ 
\begin{equation}
R(A) \,=\, r_0 \left( \frac{1}{2} \left[ A + \left( A^{\frac{1}{3}} -1
        \right)^3 \right] \right)^{\frac{1}{3}} \, ~~~~~
r_0 \,=\, \left( \frac{3}{4 \pi \rho_0} \right)^{\frac{1}{3}} \, .
\end{equation}
}
$\rho_0$ is  the nuclear matter ground state density.
{ 
If the phase-space density at the location of each nucleon} is  
  too high (i.e. the area of the nucleus  is already occupied),
then  the location of that nucleon is rejected and a new location is 
randomly chosen.
{ 
The initial momenta of the nucleons} are  randomly chosen between 0 and
the local Thomas-Fermi-momentum:
$p_F^{max} \,=\, \hbar c
        \left( 3 \pi^2 \rho \right)^{\frac{1}{3}}$,
with $\rho$ being the corresponding local proton- or neutron-density.
{ 
One disadvantage of this type of initialization}: the initialized
nuclei  are not  in their ground-state with respect to the
Hamiltonian used for the propagation.
{ 
The parameters of the Hamiltonian} are
 tuned to the equation of state of infinite nuclear matter and
to properties of finite nuclei (such as their binding energy and
their root mean square radius).
{ 
One can use a so-called Pauli-potential  in the Hamiltonian}:
This has the advantage that the  initialized nuclei remain 
 stable
whereas in UrQMD with the conventional initialization and 
propagation without
the Pauli-potential the nuclei start evaporating single nucleons after
approximately 20 - 30 fm/c.
{ 
One drawback of this potential}:  the
kinetic momenta of the nucleons are not anymore equivalent to their
canonic momenta, i.e. the nucleons carry the correct Fermi-momentum,
but their velocity is zero.
{ 
The impact parameter of a collision} is  sampled according 
to the quadratic
measure { ($dW\sim~bdb$)}. 
At a given impact parameter the{ centers of projectile
and target are placed along the collision axis in such a manner that
a distance between surfaces of the projectile and the target is equal
to 3 $fm$}. 
{ 
The momenta of the nucleons} are  transformed into the system where
the projectile and target have equal velocities directed in opposite
directions of the axis. 
{ 
After that the time propagation starts}.
During the calculation each particle is checked at the beginning of each
time step whether it will collide within that time step.
 
 The relativistic QMD models are compared to a large sample of data in
 the publications of the authors. Up to energies of about $\sqrt s$ =
 200 GeV usually a good agreement is found. The experimental
 collaborations at RHIC compare their data quite often to RQMD results, 
One example: The PHOBOS Collaboration in their white paper
\cite{PHOBOS02a} find a good agreement to  $dN_{ch}/dy$ from RQMD for
Au--Au collisions to their data, the reason for this is the chain fusion build from
the beginning into RQMD.

The main problems of the relativistic QMD models, 
which make them difficult to apply
 for shower simulations are (i) problems with the 
energy conservation,
(ii) the missing evaporation and residual nuclei and (iii) the partly 
excessive computer running times.

\section{
 DPM and QGSM models }
{The Dual Parton Model { DPM} and the  }
{ Quark Gluon String Model { QGSM} }
{  are two models, which are largely equivalent in 
their construction, only with some
 characteristic differences. }
{The 
{ { DPM} is due to Capella, Tran Than Van and collaborators
\cite{DPMREV92}}}.
{ The most detailled Monte Carlo versions of the DPM are { PHOJET}
\cite{Engel95a,Engel95d}
for h--h and $\gamma$--h
collisions and  { DPMJET} for h--A, A--A and $\gamma$--A
collisions \cite{Roesler20002}.  }

 The { QGSM} is due to Kaidalov, Ter-Martyrosian and
collaborators \cite{Kaidalov82b}. 
 The Monte Carlo version QGSJET for h--h,
 h--A and A--A 
collisions is due to  Kalmykov, Ostapchenko and
collaborators \cite{QGSJET}. 

\subsection{
 The construction of the { PHOJET }
multichain model
}
{ We restrict us in this contribution to describe the {
PHOJET} model, which is used directly for h--h collisions and for all
elementary Glauber collisions in h--A and A--A collisions 
in  { DPMJET} . There are no
essential differences in the formulations 
of the Glauber model between different
Monte Carlo models}.

The { (soft) Born cross section of the 
supercritical pomeron} has the form
$ \sigma_s = g^2s^{\alpha(0) - 1}$.
The supercritical pomeron has $\alpha(0) > 1.$, therefore
it clearly { violates unitarity}.
According to the
Froissart bound the cross section asymptotically should not
rise faster than ${(\log s)}^2$. 

If we start to construct the full model,
which is unitarized, we should
introduce some more input Born cross sections.
Very important is the { hard cross section, which we calculate 
according to the QCD improved parton model:}
\begin{eqnarray}
\sigma^{\mbox{hard}}(s,p_\perp^{
\mbox{cutoff}}) = 
\int dx_1 dx_2 d\hat{t} 
\sum_{i,j,k,l} \frac{1}{1+\delta_{k,l}} \nonumber \\
f_{a,i}(x_1,Q^2)
f_{b,j}(x_2,Q^2) \frac{d\sigma_{i,j\rightarrow
k,l}^{\mbox{QCD}}(\hat{s},\hat{t})}{d\hat{t}}
\Theta(p_\perp-p_\perp^{\mbox{cutoff}}),
\label{hard-res}
\end{eqnarray}
where $f_{a,i}(x_1,Q^2)$ is the distribution of the parton
$i$ in $a$. 

One of the most important { difference between PHOJET/DPMJET 
and QGSJET is in the $p_{\perp}$ cutoff
 $p_\perp^{ \mbox{cutoff}}$:
  PHOJET/DPMJET use{ $p_\perp^{ \mbox{cutoff}}$ 
 rising with energy}.
 QGSJET uses{ $p_\perp^{ \mbox{cutoff}}$ constant, independent of the
 energy.
 }}

We introduce furthermore the cross sections for high--mass single and
double
diffraction $\sigma_D$ and for high--mass central 
diffraction $\sigma_C$ according to the standard expressions.
The amplitudes corresponding to the one-pomeron exchange 
 are
{ unitarized applying an  eikonal formalism}. 
In impact parameter representation, the { eikonalized
scattering amplitude}
 has the structure
\begin{equation}
a(s,B) = \frac{i}{2} 
\left(\frac{e^2}{f^2_{q\bar q}}\right)^2\;
\left( 1 - e^{-\chi(s,B)}\right)
\label{eff-amp}
\end{equation}
with the {eikonal function}
\begin{equation}
\chi (s,B)=\chi_{S}(s,B)+\chi_{H}(s,B)+\chi_{D}(s,B)+\chi_{C}(s,B).
\end{equation}
Here, $\chi_{i}(s,B)$ denotes the contributions from the different
Born graphs: (S) soft part of the pomeron and reggeon, (H) hard part 
of the pomeron
(D) triple- and loop-pomeron, (C) double-pomeron graphs.

The eikonals  $\chi_{i}(s,B)$ are defined as follows
\begin{equation}
\chi_{i}(s,B) = \frac{\sigma_i(s)}{8\pi
b_i}\mathrm{exp}[-\frac{B^2}{4b_i}].
\end{equation}
 The free
parameters are fixed by a { global fit to proton-proton 
cross sections and elastic slope parameters}.  
Once the free parameters are determined, 
the probabilities for the different final
state configurations are calculated from the discontinuity of the
elastic  scattering amplitude (optical theorem). 

The total discontinuity
can be expressed as a sum of graphs with $k_c$ soft pomeron cuts, $l_c$
hard pomeron cuts, $m_c$ triple- or loop-pomeron cuts, and $n_c$
double-pomeron cuts by applying the Abramovski-Gribov-Kancheli cutting
rules . 
In impact parameter space one gets for the { inelastic cross
section}
\begin{equation}
\sigma (k_{c},l_{c},m_{c},n_{c},s,B)=\frac{(2\chi_{S})^{k_{c}}}{k_{c}!}
\frac{(2\chi_{H})^{l_{c}}}{l_{c}!}\frac{(2\chi_{D})^{m_{c}}}{m_{c}!}
\frac{(2\chi_{C})^{n_{c}}}{n_{c}!}\exp[-2\chi (s,B)]
\label{cutpro}
\end{equation}
with
\begin{equation}
\int d^2B \sum_{k_c+l_c+m_c+n_c=1}^{\infty} \sigma
(k_{c},l_{c},m_{c},n_{c},s,B) \approx \sigma_{\mbox{tot}}
\end{equation}
where $\sigma_{\mbox{tot}}$  
denotes the { total cross section} 

In the Monte Carlo realization of the model, the different final state
configurations are sampled from Eq.~(\ref{cutpro}). 
For pomeron
cuts involving a hard scattering, the { 
complete parton kinematics and
flavors/colours are sampled according to the Parton Model}. 
For pomeron cuts without hard large
momentum transfer, the partonic interpretation of the Dual Parton Model
is used: { mesons are split into a quark-antiquark pair whereas 
baryons are approximated by a quark-diquark pair}.
{ The longitudinal momentum
fractions of the partons
are given by
Regge asymptotics} .
We give it here for an event with $n_s$ soft and $n_h(n_h \geq 1)$ hard
 cut pomerons, sea--quarks are used at the chain ends if we have more
 than one soft pomeron.
\begin{eqnarray}
\rho(x_{1},...,x_{2n_{s}},...,x_{2n_{s}+2+n_{h}}) \sim
\frac{1}{\sqrt{x_{1}}} 
(\prod_{i=3}^{2n_{s}+2}
\frac{1}{x_{i}})x_{2}^{1.5} \nonumber \\
\prod_{i=2n_{s}+3}^{2n_{s}+2+n_{h}}
g(x_{i},Q_{i}) \delta(1-\sum_{i=1}^{2n_{s}+2+n_{h}}x_{i}).
\end{eqnarray}
The distributions $g(x_{i},Q_{i})$ are the distribution functions
of the partons engaged in the hard scattering. 
The momentum fractions of the constituents at the ends of the different
chains are sampled from this exclusive parton distribution, 

After all  this we have all chains defined and PHOJET/DPMJET
continues with
hadronizing all multiple chains  using the { 
Lund code JETSET (PYTHIA)}.

\begin{figure}
\includegraphics[width=6cm]{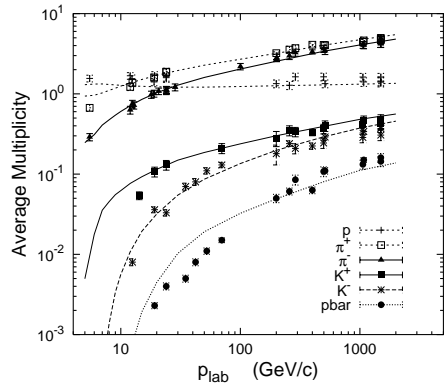}
  \caption{Average particle multiplicity proton-proton interactions.
        \textsc{Phojet}
           results (curves) are compared to experimental data
	  (symbols).}
\label{fig:muliplicity}	  
\end{figure}

\begin{figure}
\hspace*{0.0cm}\includegraphics[width=6cm]{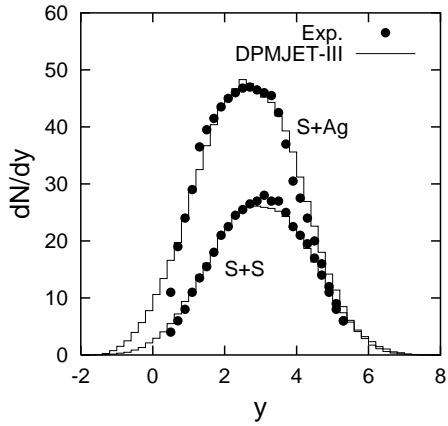}  
  \caption{Rapidity distributions of negative hadrons in central nuclear
  collisions
     at 200 GeV/nucleon.}
\label{fig:sssag}		   
\end{figure}

Now we are able to compare the multichain model  {\sc Phojet}
  with particle production data.
%
 There are many comparisons to data published in the PHOJET, DPMJET and
 QGSJET 
 literature. Here we present only two examples: the average multiplicity
 of all kinds of secondary particles in p--p collisions as function of
 the energy in Fig.\ref{fig:muliplicity} and 
 the rapidity distribution of charged hadrons in central
 S-S and S--Ag collisions in Fig.\ref{fig:sssag} at SPS energies.

\begin{figure}
\includegraphics[width=6cm]{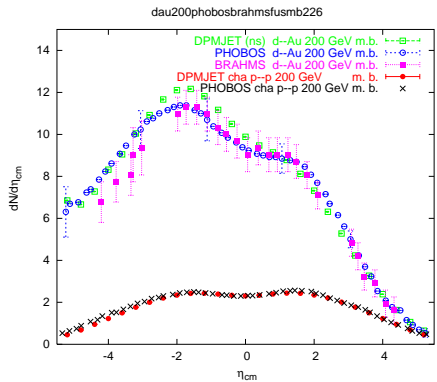}
  \caption{Pseudorapidity distribution of charged hadrons produced in
       minimum bias $\sqrt{s} = 200$ GeV d--Au and p--p collisions.
      The results     of \textsc{Dpmjet} are compared
           to  experimental data from the BRAHMS--Collaboration and
	       the PHOBOS--Collaboration. At some pseudorapidity
	      values the systematic
	           PHOBOS--errors  are given.}
\label{fig:dau}		   
\end{figure}

\begin{figure}
\includegraphics[width=6cm]{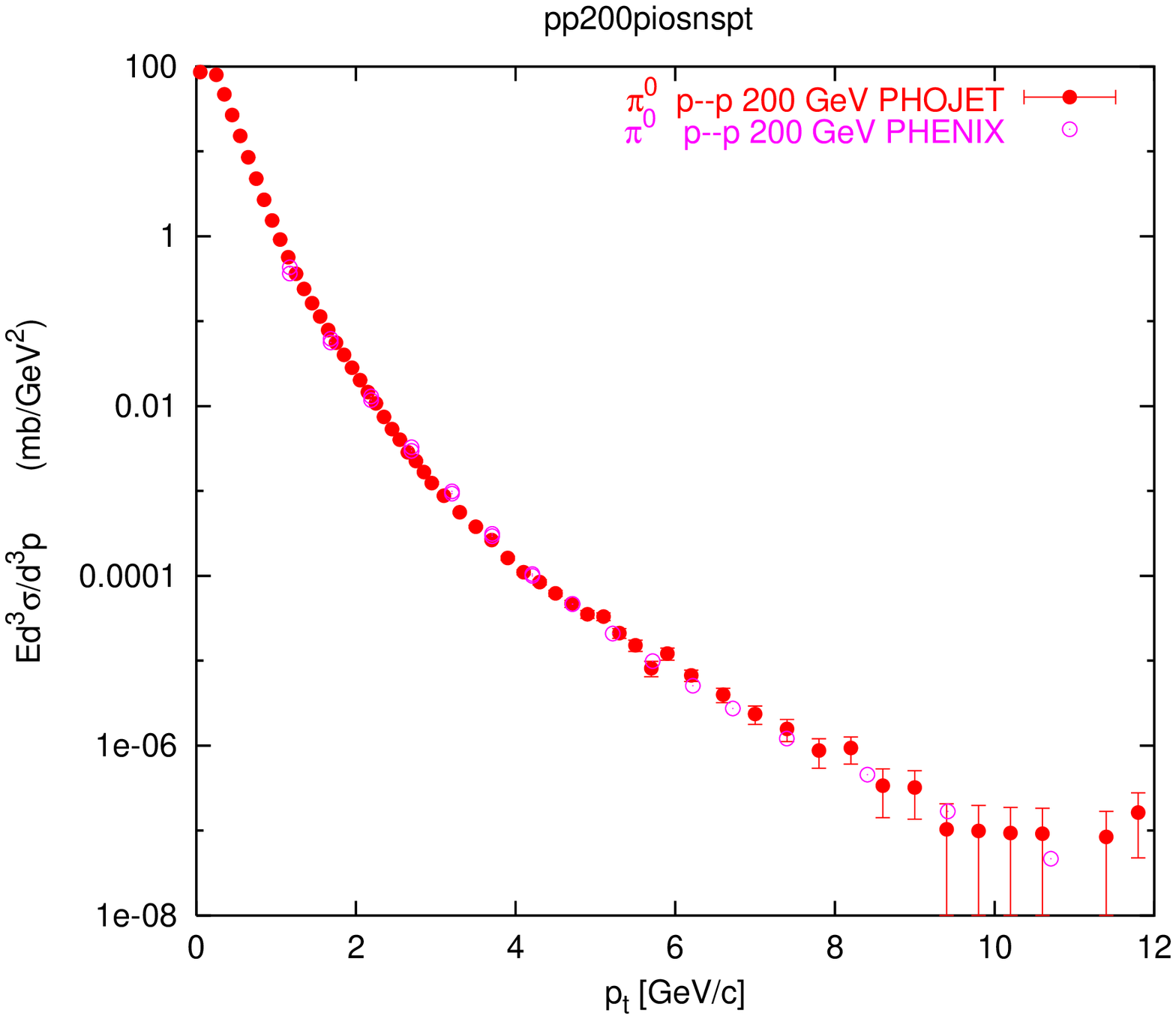}
  \caption{Transverse momentum distribution as measured in p--p
  collisions
       at $\sqrt s$ = 200 GeV by the PHENIX collaboration
           at RHIC compared to the calculation by {\sc phojet}}
\label{fig:pppt200}	   
\end{figure}

{ Next we present comparisons of the DPM--models (DPMJET--III)
with RHIC data  }
{  We first present some comparisons, where DPMJET-III 
is used in its pre--RHIC form.} In Fig.\ref{fig:dau} we compare rapidity
distributions of charged hadrons in p--p and d--Au collisions according
to DPMJET with RHIC data from the PHOBOS collaboration. In
Fig.\ref{fig:pppt200} we compare the transverse momentum distribution
measured at RHIC by the PHENIX collaboration with PHOJET calculations,
we find the hard collisions very well represented in PHOJET.

{  
 For other comparisons DPMJET needs some modifications to get agreement
with the RHIC data.} One of the most important modification is the
{\bf  Percolation of hadronic strings in {\sc Dpmjet}--III}
     
{ Using the original {\sc Dpmjet}--III with enhanced 
baryon stopping and a
centrality of 0 to 5 \% the DPMJET multiplicities  are larger than the
ones measured in
Au--Au collisions at RHIC.}
A new mechanism needed to
reduce  $N_{ch}$  and  $dN_{ch}/d\eta |_{\eta=0}$ in situations with a
produced very dense hadronic system.
     We consider only the percolation and fusion 
of {\bf soft} chains
(the transverse momenta of both chain ends are  below a  cut--off
$p_{\perp}^{fusion}$ = 2 GeV/c).
{ The condition of percolation is, that the chains overlap
in transverse space.}
We calculate the transverse distance of the chains L
and K $R_{L-K}$ 
and allow fusion of the chains for   
$R_{L-K} \leq
R^{fusion}$ = 0.75 fm. 
The chains in {\sc Dpmjet} are fragmented using the Lund code.  
Only the
fragmentation of color triplet--antitriplet chains is available in
{\sc Jetset}, however fusing two arbitrary chains could result in chains
with other colors.
Therefore, we select only chains for
fusion, which again result in  triplet--antitriplet chains. Examples
are:

{ 
(i)A { $q_1-\bar q_2$} plus a { $q_3-\bar q_4$} chain become a
{ $q_1q_3-\bar
q_2 \bar q_4$} chain.

(ii)A { $q_1-q_2q_3$} plus a { $q_4-\bar q_2$} chain become a
{ $q_1q_4-q_3$}
chain.

(iii)A { $q_3-q_1q_2$} plus a { $q_4-\bar q_1$} plus a
{ $\bar q_3-q_5$} chain
become a { $q_4-q_2q_5$} chain.

(iv)A { $q_4-\bar q_1$} plus a { $q_5-\bar q_3$} plus a
{ $\bar q_5-q_1$} chain
become a { $q_4-\bar q_3$} chain.

The expected results of these transformations are a decrease of the
number of chains.
Even when the fused chains have a higher energy than
the original chains, the result will be a { decrease of the hadron 
multiplicity $N_{hadrons}$}.
In reaction (i) we observe { new diquark and
anti--diquark chain ends}. In the fragmentation of these chains we expect
{ baryon--antibaryon production anywhere in the rapidity region of the
collision}.
Therefore, (i) helps to shift the antibaryon to baryon ratio
of the model into the direction as observed in the RHIC experiments.

In Fig.\ref{fig:auau200} we compare the pseudorapidity distributions in
Au--Au collisions  at 200*A GeV as
measured by the PHOBOS collaboration at RHIC for different centralities
with the DPMJET--III results obtained with the model including chain
percolation and fusion.

\begin{figure}
\includegraphics[height=6cm,width=8cm]{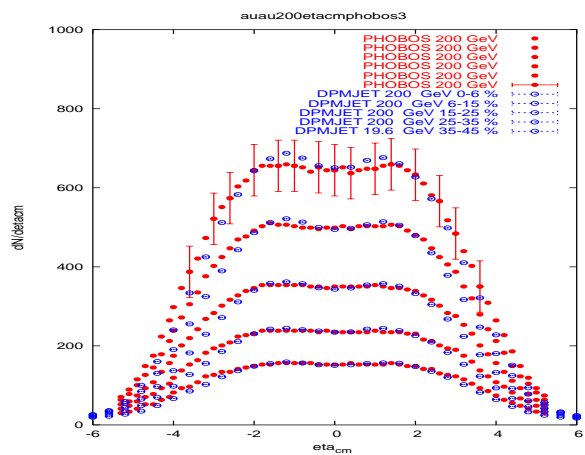}
  \caption{Pseudorapidity distributions of charged hadrons in Au--Au
       collisions at $\sqrt (s)$= 200 GeV for centralities 0--5 \%
      up to
           40--50 \%. The points with rather small error bars are
	  from the
	        \protect{\sc Dpmjet}--III Monte Carlo with chain
	      fusion as described in     the text. The data points
	      are from the PHOBOS
	           Collaboration.}
\label{fig:auau200}		   
\end{figure}

{ Further RHIC related improvements (not treated here because of the
limited space)
in DPMJET include:}

{(i)   Modified $p_{\perp}$ distributions in PYTHIA: For the
fragmentation of soft chains the Gaussian transverse momentum
distributions in PYTHIA have to be replaced by exponential ones.}

{(ii)  Collision scaling in h--A collisions: To obtain collision scaling
in DPMJET we have to change the sampling of hard chains.}

{(iii) Anomalous baryon stopping}: New diagrams lead to more
 baryons in central region. 

 {(iv) Modified diquark fragmentation}: Find missing diagram in
 diquark fragmentation, to get Antihyperon to Hyperon ratios into
 agreement with experiment.
 }

\section{ Model comparisons }
{
{\bf This section is based on a talk of  D.Heck, Karlsruhe: Comparison of models in the CORSIKA 
Cosmic Ray cascade code }} at the 
{  VIHKOS CORSIKA School 2005, Lauterbad, Germany,
May 31 -- June 5, 2005} and 
{Dieter Heck,} { private communication.}

{
{  \bf High energy models in CORSIKA used for $E_{lab} \ge $ 80 GeV}} 
include:
{  DPMJET 2.55} {(J.Ranft,Phys.Rev.D51 (1995) 64)},
{  NEXUS 2/3} {(J.Drescher et al., Phys.Rep.350 (2001) 93)},
{  QGSJET 01/II/III} {( S.Ostapchenko, Nucl.Phys.B(Proc.Suppl.)2005)},
{  SIBYLL 2.1} {(R.Engel et al.,Proc.26${}^{th}$ ICRC 1(1999)415) }.

{  \bf Low energy models in CORSIKA used for $E_{lab} \le $ 80 GeV} 
include:
{  FLUKA 2003 (only hadron production model) {(A.Fasso et al.,Proc.Monte
Carlo 2000 (2001)955)}},
{  GHEISHA 2002 {(H.Fesefeld, PITHA--85/02 Aachen (1985))},
{  UrQMD 1.3} {(S.A.Bass et al.,Prog.Part.Nucl.Phys.41 (1998)225)}.

In Fig.\ref{fig:nav} we compare the average charged multiplicity in p--p
collisions as function of the energy between the high energy models.
Only the three QGSJET models differ strongly from the rest, at high
energies the multiplicity increases much stronger than in the other
models. This is the result of the energy independent $p_{\perp}$
cut--off. In Fig.\ref{fig:xmax} we compare the $X_{max}$ of proton and
iron induced vertical showers according to the high energy models with
$X_{max}$ data. Such a comparison is hoped to determine finally the
composition of the highest energy particles in the cosmic radiation.
Even, when the models differ considerably like in Fig.\ref{fig:nav} in
their properties, we find here only modest differences in the $X_{max}$
predictions.

In Table 1 we compare the CPU times of the high energy and low energy
models. We find only NEXUS and UrQMD to need much more running time than
the other models.

\begin{figure}
\includegraphics[width=7cm]{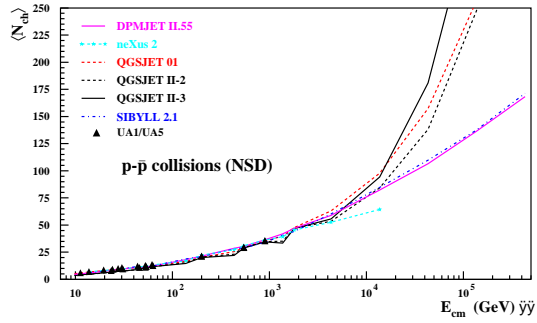}
  \caption{Table 1 Charged particle average multiplicity in $p$--$\bar p$
     collisions.}
\label{fig:nav}     
\end{figure}

\begin{figure}
\includegraphics[width=7cm]{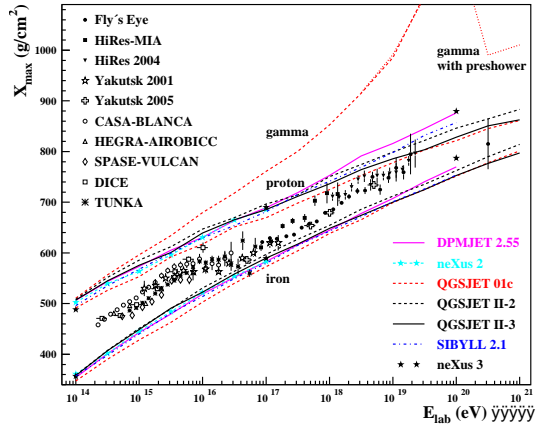}
  \caption{ Penetration depth $X_{max}$ for  { gamma},
       { proton} and { iron} induced vertical showers as
           function of the energy.}
\label{fig:xmax}	   
\end{figure}

{\small \begintable
   {\bf  CPU--times (sec)}|{\bf for DEC--Alpha 1000XP }| |\cr 
{low energy model }| { 100000 p-Air coll 10 GeV}|{high energy model }| { 10000 p-Air coll 1 PeV} \cr 
FLUKA | 181 |DPMJET 2.55 | 271 \cr
GHEISHA 2002 | 108 | NEXUS 2 | 3145 \cr
UrQMD 1.3 | 12200 |QGSJET II | 693 \cr 
 |  |SIBYLL 2.1 | 186 \endtable}

\section{ Summary and conclusions    }
{ \bf  Code comparisons}
{
\begin{description}
\item * Within 10 years of CORSIKA code comparisons: models have much
improved.
\item * Accelerator physics oriented code comparisons could help in a
similar way.
\item * include evaporation particles and residual nuclei.
\item * compare also hadron calorimeter performance, produced and
residual radioactivity.
\end{description}
}
{\bf Relativistic  QMD models}
{ 
\begin{description}
\item * Impressive performance for nucleus--nucleus collisions up to
RHIC energies.
\item * Missing: exact energy conservation, excited residual nuclei and
evaporation, residual nuclei.
( Patches to include this into FLUKA).
 {  Computer running times of these models excessively long}.
\item * Construct improved relativistic model which includes all
properties needed for cascades at accelerators, this could become a
genuine alternative to DPM, QGSM models.
\end{description}
}
{\bf   DPM, QGSM models}
{ 
\begin{description}
\item * Impressive performance for hadron--hadron, hadron--nucleus,
nucleus--nucleus, photon--hadron and photon--nucleus collisions up to
present collider energies. 
\item * Improvements through CORSIKA code comparisons.
\item * Acceptable agreement of all models up to Auger Cosmic Ray
energies.
 {  Includes also predictions for all cross sections}.
\item * These are the models which include best evaporation and residual
nuclei needed for accelerator applications. 
\end{description}
}

\begin{theacknowledgments}
I thank very much Dr. Dieter Heck from Karlsruhe for the permission to
present material from his code comparison in this talk.
\end{theacknowledgments}


\bibliographystyle{aipproc}   

\bibliography{dpm11}

\end{document}